\documentclass[conference]{IEEEtran}
\IEEEoverridecommandlockouts
\usepackage{cite}
\usepackage{amsmath,amssymb,amsfonts}
\usepackage{algorithmic}
\usepackage{graphicx}
\usepackage{subcaption}
\usepackage{textcomp}
\usepackage{xcolor}
\usepackage{multirow}
\usepackage{hyperref} 
\usepackage{balance} 
\usepackage[flushleft]{threeparttable}
\def\BibTeX{{\rm B\kern-.05em{\sc i\kern-.025em b}\kern-.08em
    T\kern-.1667em\lower.7ex\hbox{E}\kern-.125emX}}
\begin{document}

\title{The Equivalence of Fast Algorithms for Convolution, Parallel FIR Filters, Polynomial Modular Multiplication, and Pointwise Multiplication in DFT/NTT Domain\\
\thanks{This research was supported in part by the National Science Foundation
under grant number CCF-2243053.}
}

\author{\IEEEauthorblockN{Keshab K. Parhi, {\em Life Fellow, IEEE}}
\IEEEauthorblockA{\textit{Department of Electrical and Computer Engineering} \\
\textit{University of Minnesota, Minneapolis, MN 55455, USA} \\
Email: parhi@umn.edu}
}

\maketitle

\begin{abstract}
    Fast time-domain algorithms have been developed in signal processing applications to reduce the multiplication complexity. For example, fast convolution structures using Cook-Toom and Winograd algorithms are well understood. Short length fast convolutions can be iterated to obtain fast convolution structures for long lengths. In this paper, we show that well known fast convolution structures form the basis for design of fast algorithms in four other problem domains: fast parallel filters, fast polynomial modular multiplication, and fast pointwise multiplication in the DFT and NTT domains. Fast polynomial modular multiplication and fast pointwise multiplication problems are important for cryptosystem applications such as post-quantum cryptography and homomorphic encryption. By establishing the equivalence of these problems, we show that a fast structure from one domain can be used to design a fast structure for another domain. This understanding is important as there are many well known solutions for fast convolution that can be used in other signal processing and cryptosystem applications.

\end{abstract}
\section{Introduction}
Convolution of two $N$-point sequences requires $N ^ 2 $ multiplications. Fast convolution algorithms reduce the number of multiplications; thus the name \textit{fast} \cite{blahut2010fast}. Fast convolution algorithms for short lengths can be iterated to construct fast convolution algorithms for longer lengths \cite{mou1987fast, mou1991short}. Parallel finite impulse response (FIR) filters require linear increase in the number of multiplications with respect to the level of parallelism. However, starting with polyphase decomposition \cite{vaidyanathan2006multirate}, fast parallel FIR filters can be designed that require sublinear complexity \cite{parhi1999vlsi}. A fast convolution algorithm can form the basis for a fast parallel FIR filter. For example, a $L$-by-$L$ fast convolution algorithm can be used to design a $L$-parallel FIR filter in two ways. The last $(L-1)$ outputs of the $(2L-1)$-point convolution result can be added to the first $(L-1)$ outputs to design the $L$-parallel filter \cite{cheng2004hardware}. In another approach, the fast convolution structure can be transposed to obtain a fast parallel FIR filter. We do not discuss this approach in this paper; however, the reader is referred to \cite{parhi1999vlsi}. Cryptosystem applications such as post-quantum cryptography (PQC) and homomorphic encryption (HE) require computation of polynomial modular multiplication. A fast structure similar to the fast parallel FIR filter can be used to compute this product for PQC applications using less complexity in the time domain \cite{Tan_TC}. We can compute linear convolution of long sequences using FFT. The two FFTs are multiplied and then their inverse FFT (IFFT) is computed. In this paper, we describe \textit{fast pointwise multiplication} where the product of the two FFTs can be computed using fewer multiplications. In this approach, we compute the FFTs of the even and odd sequences and compute the product of the FFT using a structure similar to the fast parallel FIR filter. For HE applications, where the length of the polynomial is long, the result can be computed using number theoretic transform (NTT) and inverse NTT (iNTT). Similar to fast pointwise multiplication in the DFT domain, we can exploit fast pointwise multiplication in the NTT domain, using structures similar to fast convolution or fast parallel filter \cite{tan2023kybermat}. 

\begin{figure*}[htbp!]
\centering
\begin{subfigure}{0.4\textwidth}
    \includegraphics[width=1\textwidth]{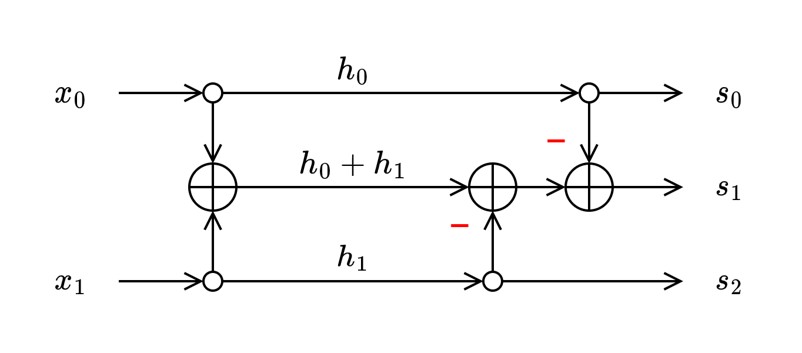}
    \caption{Fast $2$-by-$2$ convolution}
    \label{fig:FastConv}
\end{subfigure}
\begin{subfigure}{0.4\textwidth}
    \includegraphics[width=1\textwidth]{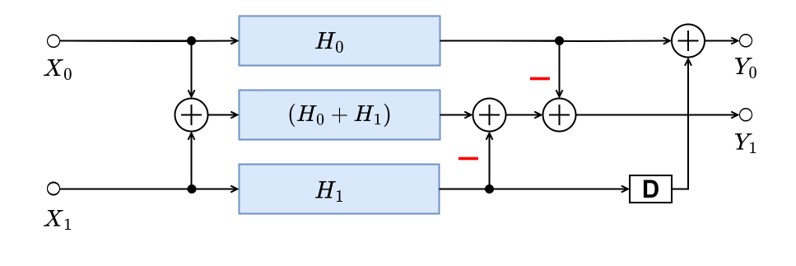}
    \caption{A $2$-parallel fast FIR filter}
    \label{fig:FastFilter}
\end{subfigure}

\begin{subfigure}{0.4\textwidth}
    \includegraphics[width=1\textwidth]{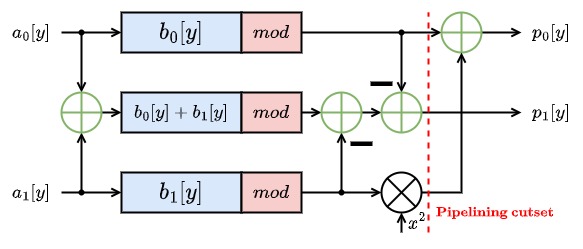}
    \caption{Fast $2$-parallel polynomial modular multiplication}
    \label{fig:FastPMM}
\end{subfigure}
\begin{subfigure}{0.35\textwidth}
    \includegraphics[width=1\textwidth]{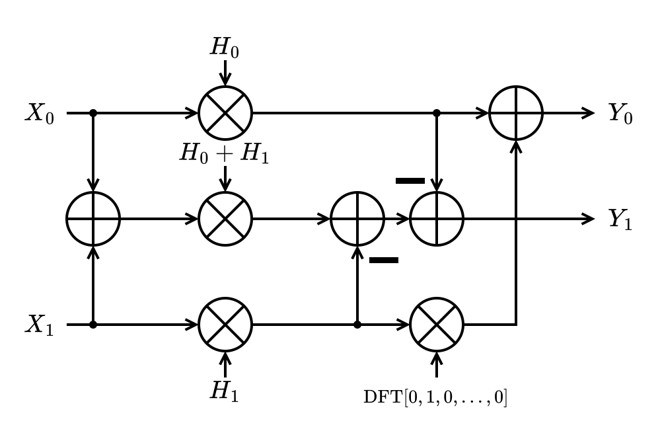}
    \caption{Fast $2$-parallel pointwise multiplication in the DFT domain}
    \label{fig:FastPM}
\end{subfigure}
\begin{subfigure}{0.5\textwidth}
    \includegraphics[width=1\textwidth]{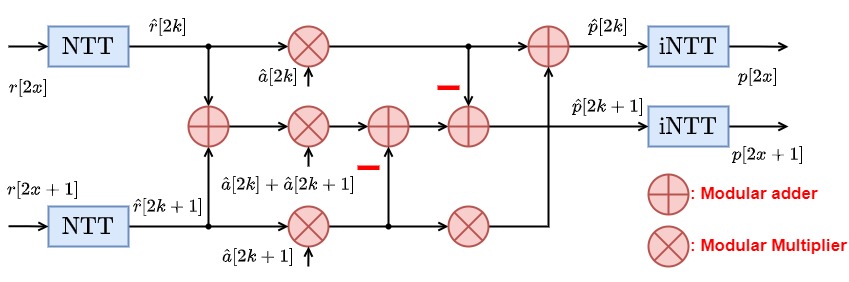}
    \caption{Fast $2$-parallel pointwise modular multiplication in the NTT domain}
    \label{fig:FastPNTT}
\end{subfigure}
\caption{The equivalence of fast structures for five problem domains.}
\label{fig:equiv}
\vspace{-1.5ex}
\end{figure*}

The goal of this paper is to unravel the equivalence of these problems, and to illustrate that a fast algorithm from one domain can form a basis for another problem domain.
Fig. \ref{fig:equiv} shows examples of fast structures for these five problem domains. It can be observed from Fig. \ref{fig:equiv} and is shown through derivations in the paper that a delay element in the fast parallel filter can be replaced by multiplication by $x^2$ in the polynomial modular multiplication, and by multiplication with DFT or NTT of a $N/2$-point sequence \{0,1,0,\dots,0\} for fast pointwise multiplication in the DFT/NTT domain.

\section{Fast Convolution}

\begin{figure}[htbp!]
\centering
\resizebox{0.35\textwidth}{!}{
\includegraphics{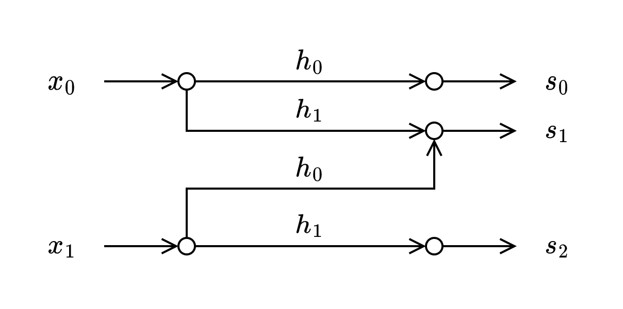}}
\caption{A $2$-by-$2$ convolution structure.}
\label{fig:Conv}
\end{figure}

Consider the convolution of two $N$-point sequences $h[n]$ and $x[n]$. A straightforward computation requires $N ^ 2 $ multiplications. Consider the 2-point sequences: \textbf{$h_2 ^ T$}=[$h_0~h_1$] and \textbf{$x_2 ^T$}=[$x_0~x_1$]. The 3-point convolution sum is described by: \textbf{$s_3 ^T$}=[$s_0~s_1~s_2$], and is illustrated in Fig. \ref{fig:Conv}. The fast convolution is described by:

\begin{equation}
    \begin{bmatrix}
            s_0 \\
            s_1 \\
            s_2 
    \end{bmatrix}
    =
    \begin{bmatrix}
     h_0 x_0 \\
     (h_0 + h_1)(x_0 + x_1) - h_0 x_0 - h_1 x_1 \\
     h_1 x_1
    \end{bmatrix}
\end{equation}
While the traditional structure requires $4$ multiplications, the fast structure, illustrated in Fig. \ref{fig:FastConv}, requires $3$ multiplications. Note that this fast algorithm is non-unique, and other fast algorithms exist for this convolution \cite{parhi1999vlsi}.

\begin{figure}[htbp!]
\centering
\resizebox{0.48\textwidth}{!}{
\includegraphics{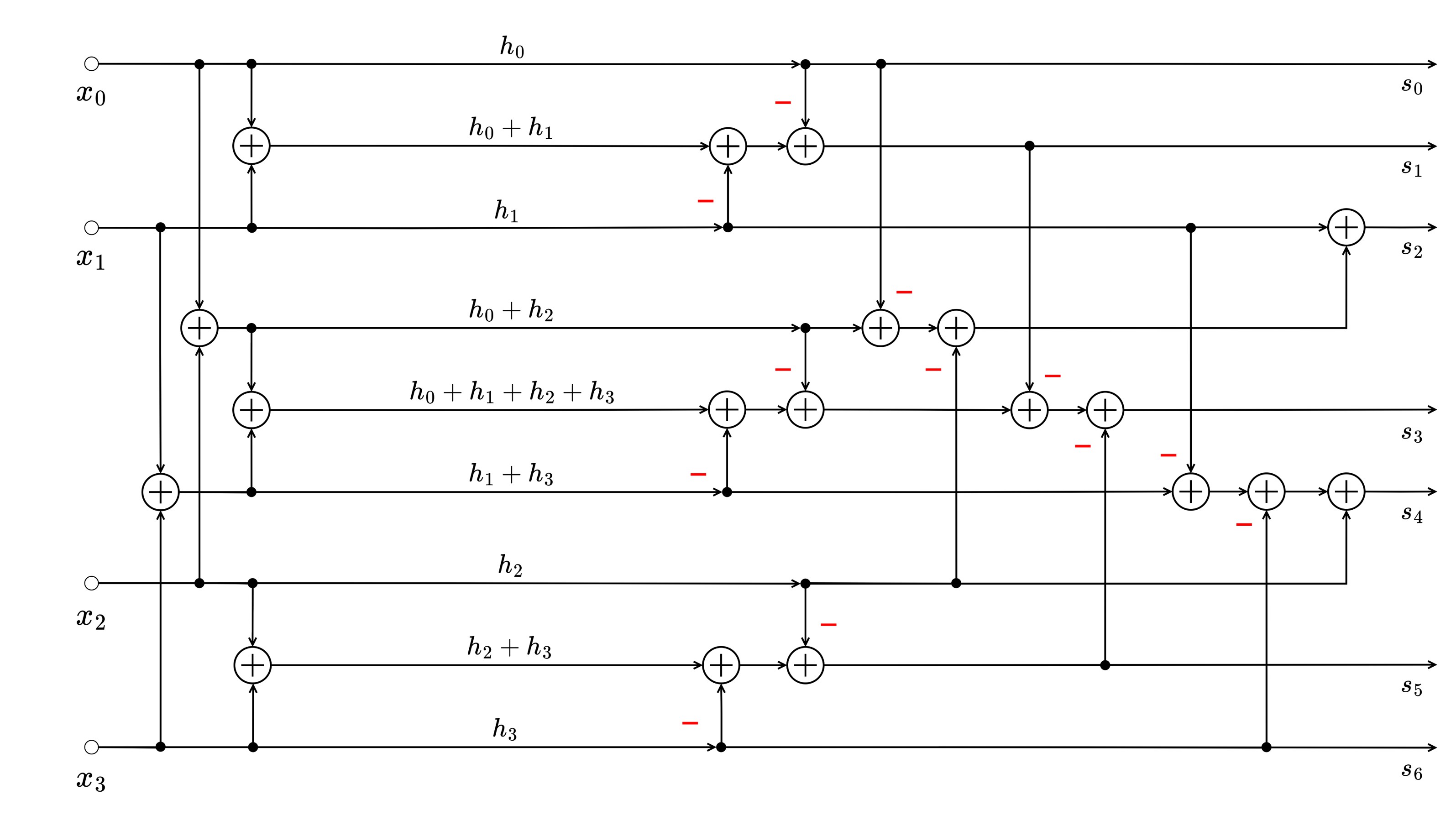}}
\caption{A $4$-by-$4$ iterated fast convolution.}
\label{fig:IFastConv44}
\end{figure}

The $2$-by-$2$ fast convolution can be written in the matrix form as:
\begin{equation}
    \mathbf{s_3}=\mathbf{Q_2}~ \text{diag}( \mathbf{P_2} \mathbf{h_2} )~ \mathbf{P_2} \mathbf{x_2} ,
\end{equation}
where
{\footnotesize
\begin{equation*}
    \mathbf{Q_2} = 
    \begin{bmatrix}
        1 & 0 & 0 \\
        -1 & 1 & -1 \\
        0 & 0 & 1
    \end{bmatrix},
    \mathbf{P_2} = 
    \begin{bmatrix}
        1 & 0 \\
        1 & 1 \\
        0 & 1
    \end{bmatrix}.
\end{equation*}
}

and $\text{diag}(\mathbf{a})$  represents the $2 \times 2$ diagonal matrix whose entries are the $2$ elements of the vector $\mathbf{a}$. The fast $2$-by-$2$ convolution algorithm can be iterated to construct a fast $4$-by-$4$ fast convolution structure, as shown in Fig. \ref{fig:IFastConv44} \cite{cheng2004hardware}.


\section{Fast Parallel FIR Filtering}
To understand the fast FIR algorithm, it is essential to first examine the formulation of parallel FIR filters using polyphase decomposition, a technique widely employed in multirate signal processing \cite{vaidyanathan2006multirate,parker1997low,parhi1999vlsi}.

An \(N\)-tap FIR filter can be expressed in the time domain as:
\begin{align}
    y[n] &= h[n] * x[n] \nonumber \\
    &= \sum_{i=0}^{N-1} h[i]x[n - i], n = 0,1,2,\dots,\infty,
\end{align}
where \( x[n] \) is an infinite-length input sequence, and the sequence \( h[n] \) contains FIR filter coefficients of length \(N\).  Similarly, in the \(z\)-domain, this can be expressed as:
\begin{equation}
    Y(z) = H(z)X(z) = \sum_{n=0}^{N-1} h[n]z^{-n} \sum_{n=0}^{\infty} x[n]z^{-n}.
\end{equation}

The input sequence \( \{ x[0], x[1], x[2], x[3], \dots \} \) can be decomposed into even-numbered and odd-numbered parts as follows:
\begin{equation}
    X(z) = x[0] + x[1]z^{-1} + x[2]z^{-2} + x[3]z^{-3} + \dots
\end{equation}
Rewriting this, we get:
\begin{align}
    X(z) &= x[0] + x[2]z^{-2} + x[4]z^{-4} + \dots \nonumber \\
    &+ z^{-1}(x[1] + x[3]z^{-2} + x[5]z^{-4} + \dots)
\end{align}
which can be simplified to:
\begin{equation}
    X(z) = X_0(z^2) + z^{-1}X_1(z^2),
\end{equation}
where \( X_0(z^2) \) and \( X_1(z^2) \) are the \(z\)-transforms of \( x[2k] \) and \( x[2k+1] \), respectively, for \( 0 \leq k < \infty \).

Similarly, the length-\(N\) filter coefficients \(H(z)\) can be decomposed as:
\begin{equation}
    H(z) = H_0(z^2) + z^{-1}H_1(z^2),
\end{equation}
where \( H_0(z^2) \) and \( H_1(z^2) \) are of length \(N/2\) and are referred to as the even and odd subfilters, respectively. For example, the even and odd subfilters of a 6-tap FIR filter \( H(z) = h[0] + h[1]z^{-1} + h[2]z^{-2} + h[3]z^{-3} + h[4]z^{-4} + h[5]z^{-5} \) are:
\begin{align}
    H_0(z) &= h[0] + h[2]z^{-2} + h[4]z^{-4}, \\
    H_1(z) &= h[1] + h[3]z^{-2} + h[5]z^{-4}.
\end{align}

The even-numbered output sequence \( y[2k] \) and the odd-numbered output sequence \( y[2k+1] \) (for \( 0 \leq k < \infty \)) can be computed as:
\begin{equation}
    Y(z) = Y_0(z^2) + z^{-1}Y_1(z^2),
\end{equation}
where:
\begin{align}
    Y_0(z^2) &= X_0(z^2)H_0(z^2) + z^{-2}X_1(z^2)H_1(z^2),  \\
    Y_1(z^2) &= X_0(z^2)H_1(z^2) + X_1(z^2)H_0(z^2). \label{eqn:2pfir}
\end{align}

\begin{figure}[htbp!]
\centering
\resizebox{0.35\textwidth}{!}{
\includegraphics{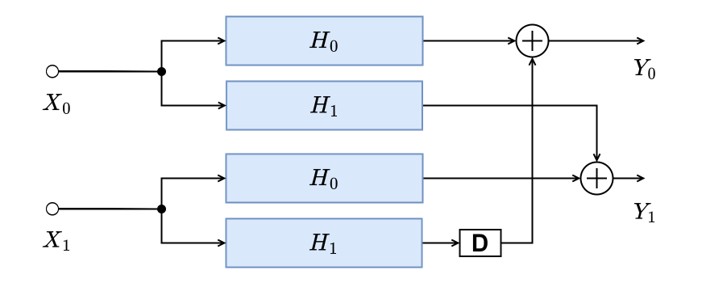}}
\caption{A $2$-parallel FIR filter.}
\label{fig:Filter}
\end{figure}

Here, \( Y_0(z^2) \) and \( Y_1(z^2) \) correspond to $Z$-transforms of \( y[2k] \) and \( y[2k+1] \) in the time domain, respectively. The filtering operation processes two inputs, \( x[2k] \) and \( x[2k+1] \), and generates two outputs, \( y[2k] \) and \( y[2k+1] \), in every iteration, as illustrated in Fig. \ref{fig:Filter}. 




The 2-parallel FIR filter in Eqn. (\ref{eqn:2pfir}) requires \( 2N \) multiplication and addition operations. Similar to the fast convolution structure, Eqn. (\ref{eqn:2pfir}) can be rewritten to obtain the fast parallel FIR filter described by:

\begin{equation}
    Y_0 = H_0 X_0 + z^{-2} H_1 X_1
\end{equation}
\begin{equation}
    Y_1 = (H_0 + H_1)(X_0 + X_1) - H_0 X_0 - H_1 X_1.
\end{equation}
and illustrated in Fig. \ref{fig:FastFilter}.
This 2-parallel fast FIR filter consists of five subfilters; however, the terms \( H_0 X_0 \) and \( H_1 X_1 \) are shared in the computation of \( Y_0 \) and \( Y_1 \). This structure requires $3N/2$ multiplications.

The fast structure is similar to the fast convolution with the only difference that the \textit{delayed} last output is added to the first output for the parallel filter. The three signals before the delay can be interpreted as the result of a fast convolution.



\begin{figure}[htbp!]
\centering
\resizebox{0.45\textwidth}{!}{
\includegraphics{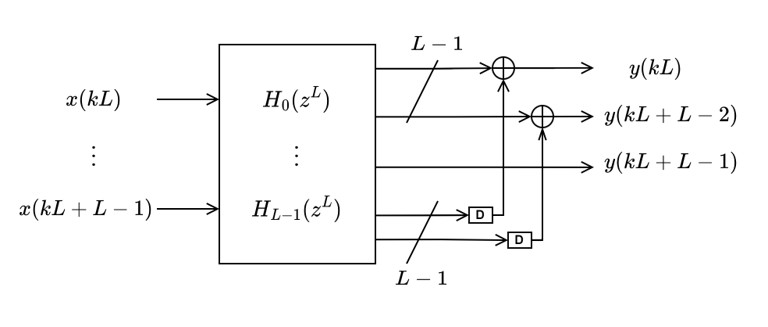}}
\caption{A $L$-parallel filter.}
\label{fig:LPFilter}
\end{figure}

\begin{figure}[htbp!]
\centering
\resizebox{0.48\textwidth}{!}{
\includegraphics{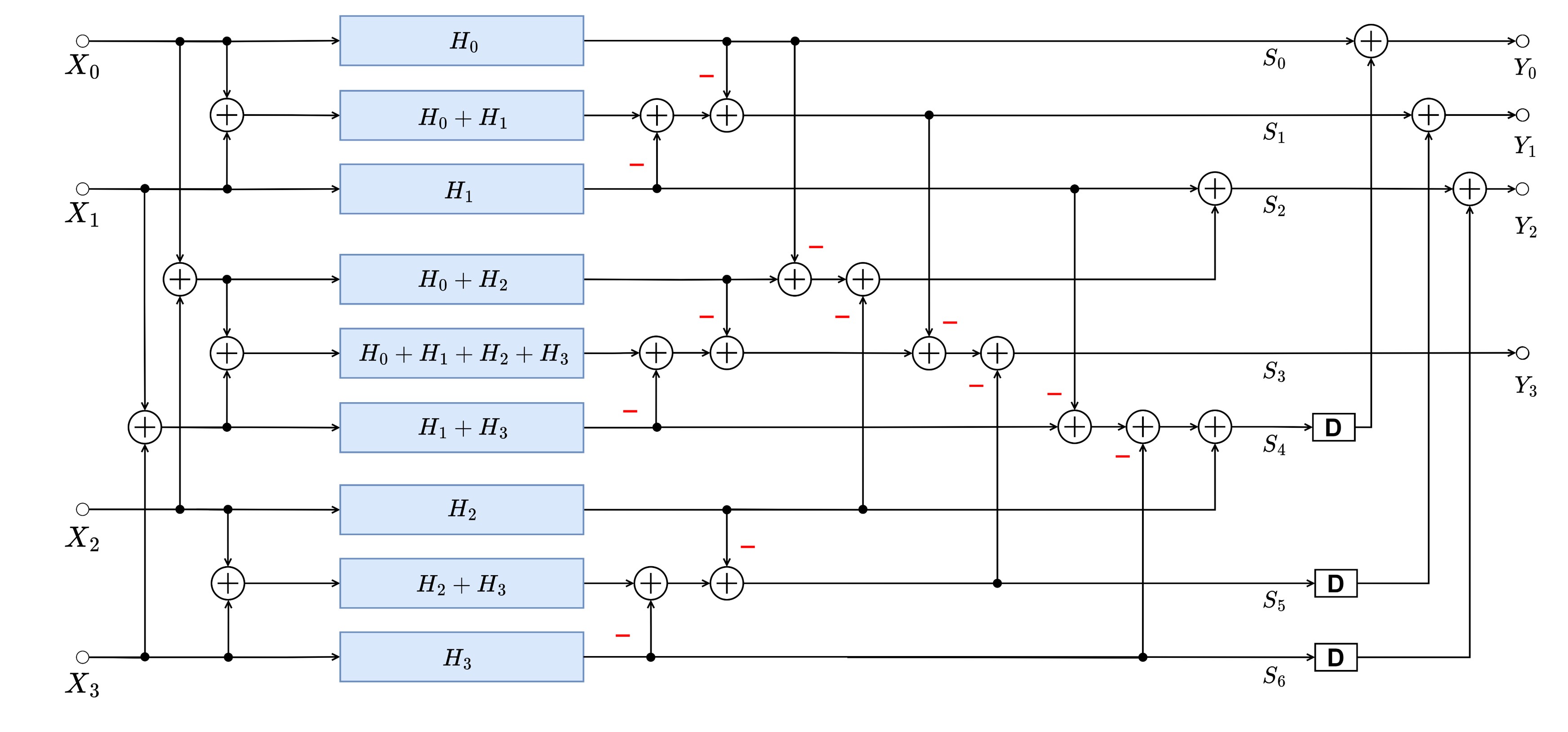}}
\caption{A $4$-parallel fast filter derived from a fast $4$-by-$4$ convolution.}
\label{fig:4PFastFilter}
\end{figure}

 In general, for a $L$-by-$L$ convolution, there are $(2L-1)$ outputs. Adding last $(L-1)$ outputs to the first $(L-1)$ leads to an $L$-parallel FIR filter as shown in Fig. \ref{fig:LPFilter}. Fig. \ref{fig:4PFastFilter} illustrates a fast $4$-parallel FIR filter derived from a fast $4$-by-$4$ convolution where the last three delayed bottom outputs are added to the top three outputs \cite{cheng2004hardware}.

\section{Fast Poynomial Modular Multiplication}

A polynomial modular multiplication is defined as:
\begin{align}
    p[x] = a[x]b[x] \mod{(x^n+1)},
\end{align}
where
\begin{align}
    a[x] &= a_0 + a_1x + a_2x^2 + \ldots + a_{n-1}x^{n-1}, \\
    b[x] &= b_0 + b_1x + b_2x^2 + \ldots + b_{n-1}x^{n-1}.
\end{align}
The polynomials $a(x)$, $b(x)$, and $p(x)$ are of length-$n$.
We can define \cite{Tan_TC}
\begin{align}
    a[x] = a_0[x^2] + xa_1[x^2], \\
    b[x] = b_0[x^2] + xb_1[x^2].
\end{align}

\begin{figure}[htbp!]
\centering
\resizebox{0.35\textwidth}{!}{
\includegraphics{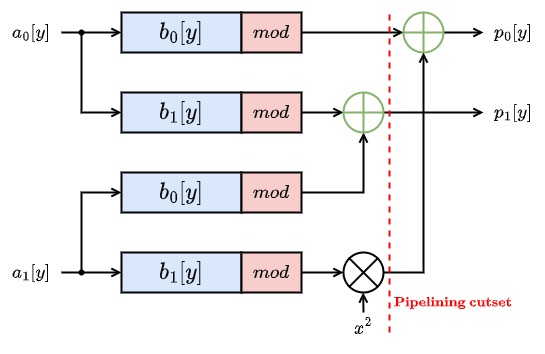}}
\caption{Polynomial modular multiplication $(y=x^2)$.}
\label{fig:PMM}
\end{figure}

Fast polynomial modular multiplication is of interest in PQC applications \cite{Tan_TC}. By applying fast algorithm similar to the $2$-parallel FIR filter, we can derive a $2$-parallel fast polynomial modular multiplier as described below:

\begin{align}
p[x] = & a[x] \cdot b[x] \bmod (x^n + 1) \\
     = & p_0[x^2] + x \cdot p_1[x^2] \\
p_0[x^2] = & a_0[x^2] \cdot b_0[x^2] + x^2 \cdot a_1[x^2] \cdot b_1[x^2] \\
p_1[x^2] = & a_0[x^2] \cdot b_1[x^2] + a_1[x^2] \cdot b_0[x^2] \\
         = & \left(a_0[x^2] + a_1[x^2]\right) \cdot \left(b_0[x^2] + b_1[x^2]\right) \nonumber \\
         & - a_0[x^2] \cdot b_0[x^2] - a_1[x^2] \cdot b_1[x^2]
\end{align}
The $2$-parallel polynomial modular multiplication is shown in Fig. \ref{fig:PMM}. The $2$-parallel fast structure is shown in Fig. \ref{fig:FastPMM}.
The main difference between fast parallel filter and the fast polynomial modular multiplication is that the delay element is replaced by multiplication by $x^2$. Note that multiplication by $x^2$ corresponds to an advance operation. So a pipeline cutset (shown as a red vertical line in Fig. \ref{fig:PMM} and Fig. \ref{fig:FastPMM}) is added to transform the structure to a causal one. The combined delay-multiplication by $x^2$ is equivalent to a switch \cite{parhi1999vlsi}. The equivalence of the fast parallel filter and the fast parallel modular multiplication was exploited to design low-latency architectures with low complexity for PQC applications \cite{Tan_TC}.

\section{Fast Pointwise Multiplication in the DFT Domain}

\begin{figure}[htbp!]
\centering
\resizebox{0.45\textwidth}{!}{
\includegraphics{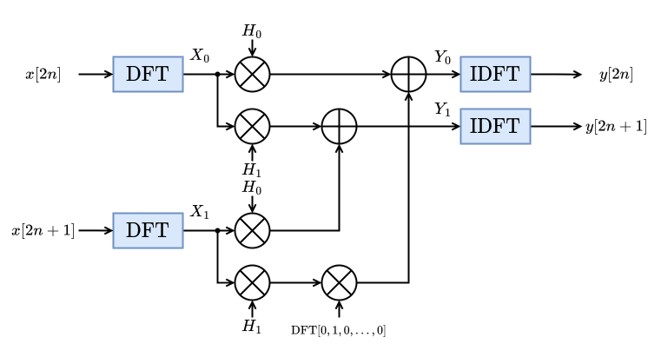}}
\caption{Pointwise multiplication in the DFT domain.}
\label{fig:PM}
\end{figure}

We can compute  $y[n] ~= ~h[n] * x[n]$ in the DFT domain using $Y[k]~=~X[k]H[k]$, where $Y[k]$ denotes the $N$-point DFT of $y[n]$. Other notations can be defined similarly. All sequences are assumed to be of length-$N$. Fast computation of $X[k]H[k]$ is referred to as fast pointwise multiplication in the DFT domain. We can express $X[k]$ in terms of the length-$N/2$ DFT of the even sequence $x[2n]$ and the odd sequence $x[2n+1]$. This is commonly referred as the merging equation in decimation-in-time FFT. Note the similarity of this representation with the polyphase representation.

\begin{align}
X[k] &= X_0[k] + e^{-j\frac{2\pi}{N}k} X_1[k] \\
H[k] &= H_0[k] + e^{-j\frac{2\pi}{N}k} H_1[k] \\
Y[k] &= X[k] H[k] \\
     &= X_0[k] H_0[k] + e^{-j\frac{2\pi k}{N/2}} X_1[k] H_1[k]  \nonumber \\
     &\quad + e^{-j\frac{2\pi k}{N}} [X_0[k] H_1[k] + X_1[k] H_0[k]] \\
Y_0[k] &= X_0[k] H_0[k] + e^{-j\frac{2\pi k}{N/2}}X_1[k] H_1[k] \\
Y_1[k] &= X_0[k] H_1[k] + X_1[k] H_0[k] \\
&= (X_0[k] + X_1[k]) \cdot (H_0[k] + H_1[k]) \nonumber \\
& - X_0[k] \cdot H_0[k] + X_1[k] \cdot H_1[k] \\
e^{-j\frac{2\pi k}{N}} &= \text{DFT}\{0,1,0,\dots,0\} \quad \text{(of a length } N/2 \text{ sequence)} \nonumber
\end{align}
Based on the above derivation, Fig. \ref{fig:PM} illustrates the pointwise multiplication in the DFT domain. This structure requires $\frac{3N}{2} \log N + N$ multiplications and $N$ additions. By applying the fast algorithm, we can obtain the 2-parallel fast pointwise multiplication shown in Fig. \ref{fig:FastPM}. Note that the main difference between the $2$-parallel fast parallel FIR filter and fast pointwise multiplication is that the delay element has been replaced by the multiplication with $\text{DFT}\{0,1,0,\dots,0\}$, a sequence of length-$N/2$.

The number of multiplications and additions are derived below. The adds inside FFT are not counted.

\begin{align*}
\#\ \text{mult} &= 6 \left( \frac{N}{4} \log \frac{N}{2} \right) + 4 \left( \frac{N}{2} \right) \\
&= \frac{3N}{2} (\log N - 1) + 2N 
= \frac{3N}{2} \log N + \frac{N}{2} \\
\#\ \text{adds} &= 5 \left( \frac{N}{2} \right)
\end{align*}

We can further investigate the number of multiplications and additions for the 4-parallel fast pointwise multiplication (not shown):

\begin{align*}
\# \ \text{Mult} &= 12 \cdot \frac{N}{8} \log \frac{N}{4} + \frac{12N}{4} \\
&= \frac{3N}{2} \cdot (\log N - 2) + \frac{12N}{4} 
&= \frac{3N \log N}{2} \\
\# \ \text{Add} &= \frac{22N}{4} + \frac{5N}{4} = \frac{27N}{4}
\end{align*}

The number of multiplications can be reduced by $N/2$ and $N$, respectively, using $2$-parallel and $4$-parallel structures. These structures can be effective in increasing the sample rate while operating the system at a lower clock rate due to the use of parallelism. We believe the connection between fast filter and fast pointwise multiplication in the DFT domain has not been established before.

\section{Fast Pointwise Modular Multiplication in the NTT Domain}

\begin{figure}[htbp!]
\centering
\resizebox{0.45\textwidth}{!}{
\includegraphics{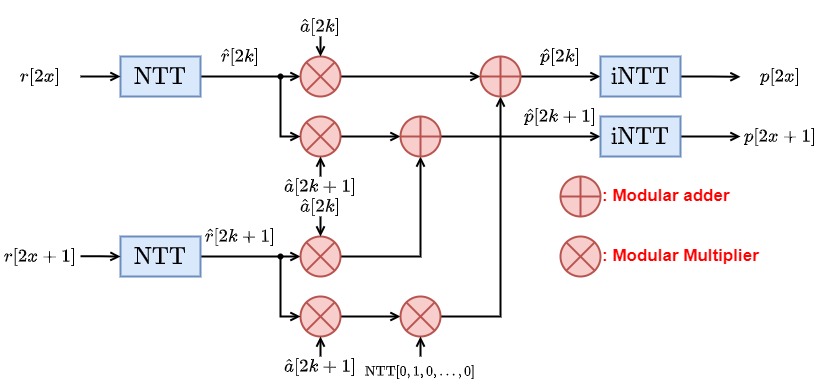}}
\caption{Pointwise modular multiplication in the NTT domain.}
\label{fig:PNTT}
\end{figure}

Consider the modular multiplication:
\begin{align}
    p[x] = a[x]r[x] \mod{(x^n+1)},
\end{align}
This can be computed using a fast architecture similar to fast pointwise multiplication in the DFT domain. In a $2$-parallel architecture, the even and odd samples of the sequence $r[x]$ can be sampled in parallel. Their $n/2$-point NTT can be computed separately. A $2$-parallel structure for the pointwise multiplication in the NTT domain to solve this problem is shown in Fig. \ref{fig:PNTT}. The main difference between the pointwise multiplication problems in the DFT and NTT domains is that the multiplication by the DFT of the $N/2$-point sequence in the DFT domain is replaced by the $N/2$-point NTT of the same sequence in the NTT domain. A fast $2$-parallel structure for the fast pointwise multiplication in the NTT domain is shown in Fig. \ref{fig:FastPNTT}. The derivation of this structure has been omitted due to lack of space. Similar structures with higher-level parallelism can be derived based on fast parallel FIR filters. This equivalence was exploited to design low-complexity structures for matrix-vector polynomial modular multiplication for the CRYSTALS-Kyber post-quantum cryptography (PQC) key-encapsulation mechanism (KEM) scheme \cite{tan2023kybermat}. This approach could also be used to reduce the hardware complexity in polynomial modular multiplication in homomorphic encryption.

\section{Conclusion}
In this paper, we have shown that the fast structures for five problem domains share similarities. These problem domains include: convolution, fast parallel FIR filter, fast polynomial modular multiplication, and fast pointwise multiplication in the DFT/NTT domain. Fast polynomial modular multiplication is useful for PQC applications. Fast pointwise multiplication in the DFT domain can be used to compute convolution of long sequences where use of parallelism can lead to high sample rates with lower clock rates. Fast pointwise multiplication in the NTT domain can be used in fast matrix-vector polynomials for PQC and for homomorphic encryption applications. Establishing the equivalence of the five problem domains means fast algorithms from one domain can be reused in another domain, thus eliminating the need to reinvent the same algorithms.

It is well known that any fast parallel filter structure can be transposed to derive another equivalent fast parallel filter. In this case, the delay elements will be in the left side of the architecture, rather than on the right side. In a fast transpose structure in the DFT/NTT domain,  the multiplication operation by the DFT or NTT of the $N/2$-point sequence will be on the left side of the structure, instead of the right side.
\section{Acknowledgment}
The author is grateful to Sin-Wei Chiu for his help in preparation of this paper. The author also thanks Weihang Tan and Yingjie Lao for helpful discussions.

\bibliographystyle{IEEEtran}
\bibliography{refs}

\end{document}